# Evolution of the specific-heat anomaly of the high-temperature superconductor $YBa_2Cu_3O_7$ under influence of doping through application of pressure up to 10 GPa.

Rolf Lortz[1], Alain Junod[1], Didier Jaccard[1], Yuxing Wang[1], Christoph Meingast[2], Takahiko Masui[3]*, and Setsuko Tajima[3]*

[1] Department of Condensed Matter Physics, University of Geneva, 24 Quai Ernest-Ansermet, CH-1211 Geneva 4, Switzerland

[2] Forschungszentrum Karlsruhe, Institut für Festkörperphysik, 76021 Karlsruhe, Germany.

[3] Superconductivity Research Laboratory-ISTEC, 10-13 Shinonome I-Chome, Koto-ku, Tokyo 135, Japan

The evolution of the specific-heat anomaly in the overdoped range of a single crystal of the high-temperature superconductor $YBa_2Cu_3O_7$ has been studied under influence of pressure up to 10 GPa, using AC calorimetry in a Bridgman-type pressure cell. We show that the specific-heat jump as well as the bulk $T_c$ are reduced with increasing pressure in accordance with a simple charge-transfer model. This new method enables us through pressure-induced charge transfer to study the doping dependence of the superconducting transition, as well as the evolution of the superconducting condensation energy on a single stoichometric sample without adding atomic disorder.

---

* Present address: Department of Physics, Osaka University, Machikaneyama-cho 1-1, Toyonaka-shi, Osaka 560-0043, Japan

# 1. Introduction

Despite a huge scientific effort, the mechanism of high-temperature superconductivity is still not understood. One approach is the investigation of the nature of phase transitions and the phase diagram of high-temperature superconductors (HTSs) by varying the charge-carrier density. The usual ways of doping HTSs is either by partial substitution of ions, or by adding and removing oxygen in the intermediate layers between the $CuO_2$ planes of the crystalline structure. In the first case, several different samples have to be used for a detailed study of the doping dependence of thermal and transport properties. The quality of the samples may then vary due to internal stress arising from ion substitution. With the second method the doping dependence on one single sample can be studied, but creation of charge carriers by oxygen doping is far from being a linear effect [1]. In addition, relaxation effects due to oxygen ordering - especially in the CuO chains of the orthorhombic compound $YBa_2Cu_3O_x$ - may influence the charge-carrier density [2-4].

The ideal case would be a mechanism to study a large doping range on one single crystal without changing stoichiometry and, thus, without adding atomic disorder. Field-effect doping [5] is believed to be a candidate, but poses significant challenges in practice and cannot be applied to bulk single crystals. In this article we present a second possibility: pressure induced doping of $YBa_2Cu_3O_x$ through charge transfer from the intermediate layers into the $CuO_2$ planes [6-10,11], in combination with AC calorimetry [12,13] in a Bridgman-type pressure cell [14,15]. This method allows us to study the phase diagram, the specific-heat anomaly at $T_c$ and the evolution of the superconducting condensation energy through a large doping range. In addition, specific heat is a thermodynamic quantity and, thus, gives access to the pressure dependence of the real bulk $T_c$. This might differ in some cases strongly from the $T_c$ derived using standard high-pressure experiments like AC susceptibility or resistivity measurements [16]. The calorimetric method has been developed to trace phase transitions of heavy-fermion compounds having a huge specific heat at low temperatures [13], while it was widely believed that it might become increasingly difficult at higher temperatures. We developed this method further to study the tiny specific-heat anomalies of HTSs at the superconducting transition (which represent typically only a fraction of 1-4 % of the total specific heat) under influence of high pressure.

In this paper we focus on the overdoped side of the phase diagram where HTSs become more and more metallic and where it is expected that thermal fluctuations of the order parameter are only of little importance [17]. We show that the value of the bulk $T_c$ is reduced with increasing pressure, while we find only minor changes in shape and size of the specific-heat anomaly. Further, we discuss the doping dependence of the superconducting condensation energy compared to data reported in literature based on doping through ion substitution.

# 2. Experiment

A high quality, detwinned overdoped $YBa_2Cu_3O_7$ single crystal was used for the present study. It shows fully reversible behavior in magnetic fields at ambient pressure, with very sharp specific-heat peaks on the first-order vortex-melting line as will be reported in a second article [18]. This is a clear indication for its high quality. It was annealed at 400 °C in 90 bar $O_2$ pressure for 200 hours to obtain nearly full oxygenation of the CuO chains. The bulk superconducting transition temperature $T_c$=88 K at ambient pressure was determined with an adiabatic calorimeter. A small piece of 0.6 mm along the b-axis and 0.25 mm along the a-axis was cut from the sample and polished down to a thickness of 20 μm along the c-axis using a diamond saw. We have chosen a detwinned crystal to limit broadening effects of $T_c$ due to the strong uniaxial pressure dependencies of opposite sign along the a- and b-axis [19,20]. A small nonhydrostaticity due to the use of a solid pressure-transmitting medium might otherwise give rise to a uniaxial pressure component, creating a pressure effect with alternating sign in the different domains of a twinned crystal. Further, we have chosen a fully oxygenated stochiometric sample to avoid pressure-induced oxygen-ordering effects in the $CuO_2$ chains, known to appear around room temperature in presence of oxygen vacancies [2-4]. We note that in our experimental setup the pressure has to be changed at room temperature.

The sample was fixed on a disc of steatite, a soft solid, serving as a pressure-transmitting medium, inside a pyrophyllite gasket which is mounted on one of two tungsten-carbide anvils of the pressure cell (see figure 1a and b). Eight 50 μm thick Au-wires (annealed) were placed across the gasket at about half of its thickness, wedged into small channels cut with a razor blade and covered with compressed pyrophyllite powder. Inside the gasket, two chromel-constantan thermocouples made of flattened 25 μm thick wires were laid on the sample together with one Au-wire of same thickness and one Pb-wire. This allows us to also perform DC four-probe resistivity measurements along the b-axis with the Au/Pb-wire as currents leads and the two thermocouples as voltage leads. The absolute value of the resistivity has been estimated by a simulation with graphite paper [21].The contacts with the sample were established due to the force of the pressure-transmitting medium after closing the pressure cell. The Pb-wire, which is connected in series with the sample, has been chosen as a manometer using the well-known pressure dependence of the superconducting transition of Pb. For the analysis, the calibration of Thomasson et al. [22] was used [23]. An estimation of the pressure gradient is given by the width of the superconducting transition in the resistivity of the Pb-manometer, which is typically ~10 % of the absolute value of the pressure.

As pressure is determined at low temperatures with this method, the thermal expansion of the parts of the pressure cell had to be compensated by the choice of materials. In an additional experiment with a Bi-sample as a manometer at room temperature [24] in comparison with a Pb-manometer at low temperatures no variation within 10 % accuracy was found. For the specific-heat measurements the resistance of one of the thermocouples was used as an AC heater, driven by the AC voltage output of a digital lock-in amplifier [25]. The second thermocouple was used to detect the resulting temperature modulation of the sample. A simple model of the AC calorimetry system [12] predicts the amplitude and phase of the temperature oscillations ($T_{ac}$) induced by AC heating:

$$T_{AC} = \frac{P_0}{K + i\omega C} \quad (1)$$

where $P_0$ is the heating power, $K$ the thermal conductivity between sample and heat bath, $C$ the heat capacity, and $\omega/2\pi$ the excitation frequency, assumed to be low enough that the thermometer can follow the temperature oscillations. The signal therefore contains a contribution from the specific heat and from the thermal coupling to the surroundings. For frequencies $\omega \gg \omega_c$, where $\omega_c$ is the cut-off frequency $K/C$, the sample contribution dominates the signal, and $|T_{AC}|$ can be considered to be inversely proportional to the heat capacity (which we assume to be dominated by the sample). For $\omega \ll \omega_c$, the signal approaches the DC limit and gives a measure of the mean elevation of the sample temperature over that of the bath. Using formula (2) [26]:

$$C = \frac{P_0}{(w_2^2 - w_1^2)^{1/2}} \left( \frac{1}{|T_{AC, w_2}|^2} - \frac{1}{|T_{AC, w_1}|^2} \right)^{1/2} \quad (2)$$

where $T_{AC,w_1}$ and $T_{AC,w_2}$ are the modulated components of the sample temperature for two applied frequencies $w_1$ and $w_2$ ($w_2 > w_1$), we can cancel the contribution of the thermal conductivity of the surroundings by repeating the measurement at different frequencies ($w/2\pi = 444$ Hz and 778 Hz). We note that this model of AC calorimetry is oversimplified for the complicated system inside the pressure cell to extract an absolute value of the specific heat. In the following we will show that the specific heat as estimated by equation (2) is nevertheless a useful quantity to study the evolution of phase transitions of high-temperature superconductors under the influence of high pressure.

## 3. Results

Figure 2 shows the specific heat $C/T$ of $YBa_2Cu_3O_7$ in arbitrary units for 1.1, 3, 5, 7.4 and 10 GPa. The presented data have been taken upon increasing pressure. To check whether the pressure effect is reversible, the pressure was released afterwards from 10 to 7 GPa. We found good consistency concerning the $T_c(p)$ value (see figure 5 for details) and the size of the anomaly [27]. Figure 2 shows clearly the anomaly at $T_c$, which is shifted to lower temperatures upon increasing pressure. The contribution of the specific-heat anomalies at $T_c$ is of the order of 1-2 % of the total signal, clearly above the noise level of about 0.07 % (r.m.s.). The value of the total signal is reduced with pressure up to 7 GPa and then increases again. This strong variation cannot be explained by a change of specific heat of the sample alone and must arise from contributions of the sample surroundings, i.e. mainly the pressure-transmitting medium. Apart from that, the increase of the total signal with increasing temperature is much stronger than expected from the specific heat at ambient pressure. At the present stage the origin of this behavior is not understood. For the further analysis of our data, we assume that these contributions are additive and can be considered as a background. Contrary to the change in the total value of specific heat, the changes in the size of the specific-heat anomaly with increasing pressure are much smaller. The anomaly broadens upon increasing pressure, most probably due to deviations from ideal hydrostatic conditions in the solid pressure-transmitting medium.

Figure 3 shows the corresponding resistivity measurements. The specific-heat anomaly appears close to the temperatures where an extrapolation of the steepest slope during the transition approaches zero (see figure 5 for details). The measurement at lowest pressure (1.1 GPa) is missing in the resistivity due to a missing contact in one of the current leads, which was established at higher pressures. In the 3 GPa measurement the resistivity is already slightly reduced below ~140 K, which might reflect the presence of superconducting fluctuations above $T_c$ [17,28-30]. This reduction is less pronounced at 5 GPa and finally disappears at higher pressures, which can be clearly seen in the inset of figure 3 where the resistivity is normalized by a linear extrapolation of the normal-state resistivity. This behavior suggests that the phase stiffness of the superconducting order parameter is increasing under the influence of pressure. The resistivity above $T_c$ is more and more reduced with increasing pressure, reflecting the increasing charge-carrier density by pressure-induced doping in accordance with literature [31].

For the further analysis of the evolution of shape and size of the specific-heat anomaly at $T_c$ under pressure, we first subtracted from the $C/T$ data a linear background with same slope for all curves [32], fitted in a small temperature range just above $T_c$ (figure 4a). Afterwards we scaled the temperature axis by choosing the reduced temperature $[T-T_c(p)]/T_c(p)$ [33], and finally normalised the data by the value of the specific-heat jump D$C/T$ (figure 4b). As the anomaly is more strongly broadened at higher pressures, we simulated a $T_c$-broadening for the data at lower pressures by smoothing the data. After this treatment, the size of the anomaly is slightly reduced upon increasing pressure, confirming results from specific-heat data of Ca-doped samples on the overdoped side of the phase diagram [34]. This suggests in accordance with our resistivity results that the main effect of the application of pressure is a doping effect.

Although our data analysis depends on the simulated boadening and on the assumption that the additional contributions from the pressure-transmitting medium are additive, we can clearly see even from the raw data that the changes in D$C$ (which is closely related to the superconducting condensation energy) are small, whereas $T_c$ is decreased by 12 degrees through the application of pressure. This behavior contrasts with that known from the underdoped side of the phase diagram of HTS: for a comparable $T_c$ change D$C$ has been reported to decrease by a factor of ~4 [34,35]. That is at least a factor of 3 more than what we find on the overdoped side. The condensation energy varies thus much less on the overdoped side, reflecting the absence of strong fluctuations [17,28-30] or the absence of a pseudogap [36] above $T_c$.

The shape of the $C/T$ anomaly at $T_c$ is hardly influenced by the treatment of our data. Apart from the broadening, no significant changes are observed, as seen in figure 4b where all curves for different pressures are scaled on top of each other. Although the transition is too broad to perform an analysis of the evolution of fluctuations close to $T_c$, it seems that the nature of the transition does not change at least in this range of the overdoped side of the phase diagram. The possible presence of fluctuations above $T_c$, indicated in the resitivity measurements, is thus only of little importance for the bulk transition.

The variation of the transition temperature $T_c(p)$ is derived from the scaling of the data in figure 4b [33]. The result is plotted in figure 5 together with values derived from the resistivity where we used an extrapolation of the steepest slope of the resistive transition to zero to obtain a r=0 value. We find that $T_c$ decreases for all pressures with an initial slope of ~ -1 K/GPa. This is in good agreement with reported pressure dependencies found from thermal-expansion [11] and was recently confirmed by AC-susceptibility data of a second small piece of the same crystal [37], taken in a He-gas pressure cell under true hydrostatic conditions. Around 5 GPa $T_c(p)$ as derived from specific heat is steeper, creating a broad step in the curve. The step is less pronounced in the data derived from resistivity, but it is difficult to extract an accurate $T_c$-value from the broad resistive transition whereas $C(p)$ scaling is less sensitive to broadening. At the moment we cannot fully exclude an artefact due to deviations from true hydrostatic conditions. A non-linear charge transfer or charge-ordering effects in the $CuO_2$ planes could give another explanation. The anomaly in the $T_c(p)$ curve is also present in the $T_c(V)$ curve, where the change in volume of the unit cell as a function of pressure $V(p)/V(p=0)$ is deduced from literature data for $YBa_2Cu_3O_7$ [10,38].

From the size of D$C/T$ it is possible to estimate roughly the evolution of the superconducting condensation energy $U_0$ with pressure (figure 6a). As our specific heat is given in arbitrary units, we used data for a Ca-doped sample with a similar $T_c$ [39] to calibrate the $U_0$ value at $p$=1.1 GPa. Under the assumption that there are only minor changes to the Sommerfeld constant $g_n$, we find that $U_0$ decreases smoothly with decreasing volume. The $T_c(p)$ values allow us to estimate the volume dependence of the charge-carrier density $n_h$ (in holes per formula unit and $CuO_2$ plane) using formula 3 given by Tallon et al. [40]:

$$T_c/T_{c,max} = 1-82.6(n_h-n_h^{optimal})^2 \qquad (3)$$

For x=6.93 YBa$_2$Cu$_3$O$_x$ is optimally doped ($n_h^{optimal}$ =0.16 [40]) and a pure charge-transfer model would predict $dT_c/dp$=0. Experimentally, an additional intrinsic component to the pressure effect on $T_c$ is observed [41] with a value of $(dT_c/dp)^{intrinsic}$= +0.52 K/Gpa [11,19] or $(dT_c/dp)^{intrinsic}$ = +0.4 K/Gpa [38] on comparable YBCO single crystals. Its origin may be traced to an increase in the coupling stength. Assuming it to be constant over the investigated doping range, and neglecting other contributions due to pressure-dependent quantities such as e.g. $n_h^{optimal}$, we use this value of the intrinsic contribution to isolate the pressure-induced change of $T_c$ due to charge transfer alone: $(dT_c/dn_h)(dn_h/dp)$, according to formula 4 [41]:

$$(dT_c/dn_h)(dn_h/dp) = dT_c/dp - (dT_c/dp)^{intrinsic} \qquad (4)$$

Using the calculated $n_h$ values, we find a pressure-induced change of the carrier density $dln(n_h)/dp$ of 1 to 2.2 %/GPa, whereas values as high as 9 % are reported from Hall-effect measurements for YBa$_2$Cu$_3$O$_7$ [42]. Using formulas 3, 4 and an intrinsic pressure effect of +0.5 K/GPa, the Hall-effect result would lead to a $T_c$ decrease of -9.5 K/GPa in the overdoped region under study. This is in contradiction to the observed pressure effect on $T_c$. Jorgensen *et al.* [43] utilized a modified version of the 'valence bond sum' technique and found a value of $dln(n_h)/dp$ ~3 to 4 %/GPa which is closer to our estimate. This discrepancy stimulates further experimental work including simultaneous Hall effect and specific-heat measurements, which can be implemented in principle in our pressure cell at the cost of minor modifications.

As the $T_c(n_h)$ curves follows an inverted parabola at least on the overdoped side, the same plateau-like anomaly is found around a value of $V(p)/V(p=0)$=0.985 in the $T_c(V)$ or $T_c(p)$ curve. In figure 6b we finally plotted the pressure dependence of $U_0$ vs. $n_h$. The behaviour thus found on the overdoped side of the phase diagram nicely confirms on a stoichometric sample without adding disorder what was known from measurements on a series of Ca-doped samples [34,39]. It can be mainly explained by a decreasing strength of the pairing interaction in a superconductor resembling more and more a standard metal - far away from the doping range where phase fluctuations [17,28-30] and the opening of a pseudogap [36] are of particular importance for its thermal properties.

The broad step-like feature in the $T_c(p)$-curve - as already discussed above - creates a plateau-like anomaly in the $U_0$ vs. $n_h$ curve around a value of $n_h$=0.194. Although only indicated by one single data point in figure 7c, a similar feature seems to be present in the $U_0(n_h)$ data of Tallon et al. [39]. This stimulates further work on the overdoped side of the phase diagram of YBCO.

## 4. Concluding remarks

Although more work has to be done to be able to interpret the total specific-heat values of AC-calorimetry data in high pressure, the present work shows that it is a useful tool to gain access to the pressure dependence of the bulk $T_c$ of high-temperature superconductors, to the condensation energy, and the doping dependence of the shape of the specific-heat anomaly, which contains information about superconducting fluctuations and the pseudogap. This stimulates further experiments on underdoped samples such as e.g. the stochiometric compound YBa$_2$Cu$_4$O$_8$, i.e. on the still somewhat mysterious side of the phase diagram of HTSs, where the interplay between phase fluctuations and the pseudogap is still far from being understood.

**Acknowledgments**

This work was supported by the Swiss National Science Foundation through the National Centre of Competence in Research "Materials with Novel Electronic Properties-MaNEP". This work was partially supported by the New Energy and Industrial Technology Development Organization (NEDO) as the Collaborative Research and Development of Fundamental Technologies for Superconductivity Applications.

[33] In addition we need a working definition of $T_c$ for one of the curves; we chose the inflection point in the data at p=1.1 GPa after subtracting a linear background fitted above $T_c$, but before we applied the broadening to the transition.

[34] J.W. Loram, K.A. Mirza, J.R. Cooper, and J.L. Tallon, *J. Phys. Chem. Solids* **59**, 2091 (1998).

[35] M. Roulin, Ph.D. Thesis (1998), *Specific Heat of the High-$T_C$ Superconductor $YBa_2Cu_3O_x$, near $T_C$ versus the Oxygen Concentration,* University of Geneva, Switzerland.

[36] T. Timusk and B. Statt, *Report on Progress in Physics* **62**, 61 (1999).

[37] T. Tomita and J.S. Schilling (private communication).

[38] W.H. Fietz, R. Quenzel, H. A. Ludwig, K. Grube, S. I. Schlachter, F. W. Hornung, T. Wolf, A. Erb, M. Kläser and G. Müller-Vogt, Physica C 270, 258 (1996).

[39] J. L. Tallon, G. V. M. Williams, and J. W. Loram, *Physica* C **338**, 9 (2000).

[40] J. L. Tallon, C. Bernhard, H. Shaked, R.L. Hitterman, and J.D. Jorgensen, *Phys. Rev.* B **51**, 12911 (1995).

[41] J. J. Neumeier and H. A. Zimmermann, Phys. Rev. B **47** (1993) 8385.

[42] C. Murayama et al. Physica C 183, (1991) 277.

[43] J.D. Jorgensen, S. Pei, P. Lightfoot, D.G. Hinks, B.W. Veal, B. Dabrowski, A.P. Paulikas, R. Kleb, and I.D. Brown, Physica C 171, (1990) 93.

[44] S. Deemyad, Ph.D. Thesis (2004), Washington University in St. Louis.

[45] B. Bireckhoven and J. Wittig, *J. Phys.* E **21**, 841 (1988).
9

**Figure 1.** (a) Assembly of the sample, thermocouples and current leads on a disc of steatite (white) inside a pyrophyllite gasket (grey) mounted on one of the tungsten-carbide anvils of the pressure cell. The steatite disc serves as a pressure transmitting medium. A second disc of steatite is subsequently put on top of this montage before closing the cell. Under pressure, the total thickness of the cell is about 180 μm.
(b) Schematic drawing showing the $YBa_2Cu_3O_7$ sample, the wires and the Pb-manometer inside the pressure cell.

**Figure 2.** $C/T$ of the $YBa_2Cu_3O_7$ sample (including some contribution from the steatite) in 1.1, 3, 5, 7.4 and 10 GPa. The kinks are the anomalies at the superconducting transition temperatures. Upper left inset: Specific heat $C/T$ at ambient pressure, measured in an adiabatic calorimeter using the full-size crystal (53 mg). Lower right inset: Electronic contribution to the specific heat $C^{electr.}/T$ at $p$=1.1 GPa after subtraction of a polynomial background fitted at 92-97 K and 65-70 K.

**Figure 3.** Resistivity along the orthorhombic b-axis of the detwinned $YBa_2Cu_3O_7$ sample in 3, 5, 7.4 and 10 GPa [21]. Inset: Resistivity normalized to a linear extrapolation of the normale-state contribution.

**Figure 4.** (a) $C/T$ data after subtraction of a linear background fitted above $T_c$. The data at 1.1 GPa and 3 GPa were broadened by averaging to compare the anomaly with the data at higher pressure (see text).
(b) Same data plotted as a function of the reduced temperature $[T-T_c(p)]/T_c(p)$. The maximum has been normalized to one. This determines the scaling factor $DC(p)/T$, and shows that the shape of the transitions remains qualitatively unchanged. Left inset: Scaling factors $DC(p)/T$ as a function of pressure normalized to the value at 1.1 GPa. Right inset: $T_c(p)$ as a function of pressure as derived from the scaling.

**Figure 5.** $T_c$ as a function of pressure (upper scale) and relative change in volume $V(p)/V(0)$ (lower scale) derived from scaling the specific heat in Figure 4b in comparison with resistivity and ac susceptibility [37] values. We used the temperature where the extrapolation of the resistive transition goes to zero (see text). The closed triangles were taken upon increasing, the open triangle upon releasing pressure [27]. The susceptibility data (χ') has been taken in a He-gas pressure cell on another small piece of the same crystal. The variation of the unit cell volume is deduced from the pressure using compressibility data obtained by X-ray diffraction [10,38].

**Figure 6.** (a) Condensation energy versus pressure (upper scale) and pressure-induced relative change in volume $V(p)/V(0)$ (lower scale). The absolute value of the condensation energy was calibrated by scaling the data in Figure 6b to that obtained from Ca-doped samples using specific-heat data from Tallon et al. [39]. Lower left inset: $T_c$ versus charge-carrier density $n_h$ (in holes per formula unit and $CuO_2$ plane) calculated using the formula given by Tallon *et al.* [40]. Only the variation of $T_c$ due to charge transfer induced by pressure is considered here, i.e. an intrinsic variation at constant doping $(dT_c/dp)^{intrinsic}$ =0.52 K/GPa has been subtracted (see text for details). Upper right inset: Pressure-induced change in the charge-carrier density $n_h$ versus relative change in volume.
(b) Condensation energy $U_0$ versus charge-carrier density $n_h$ in comparison with data from Ca-doped samples [39].

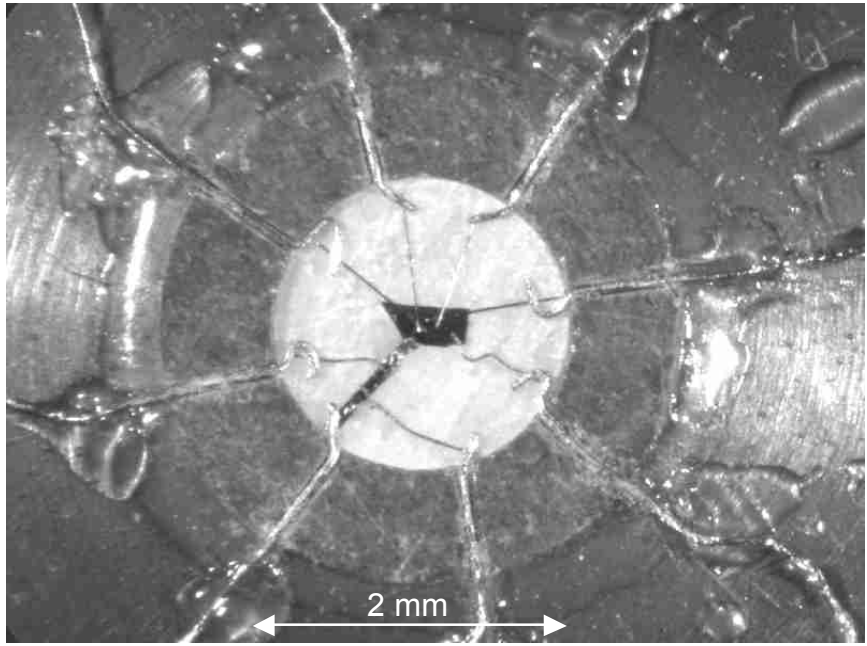

(a)

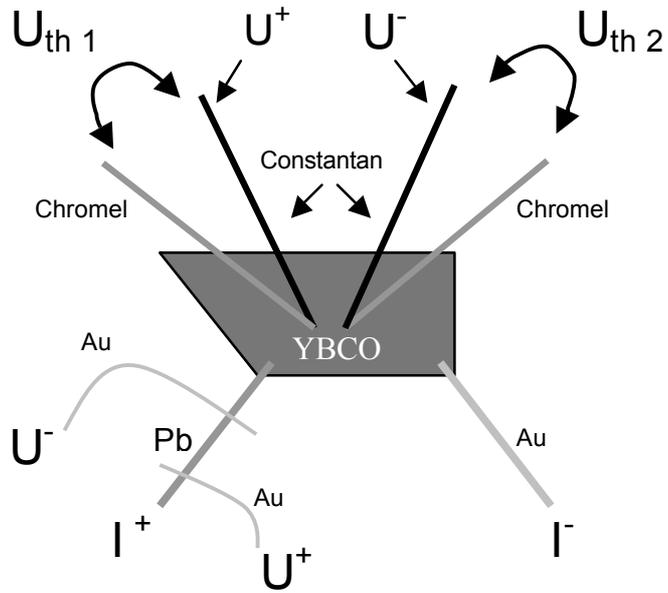

(b)

1a
1b

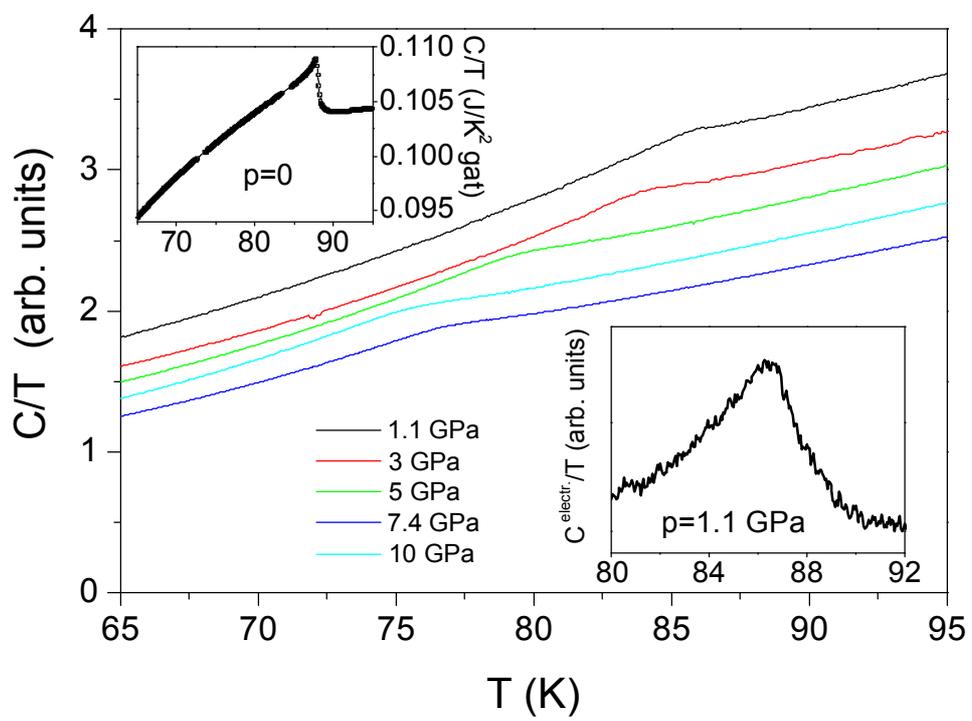

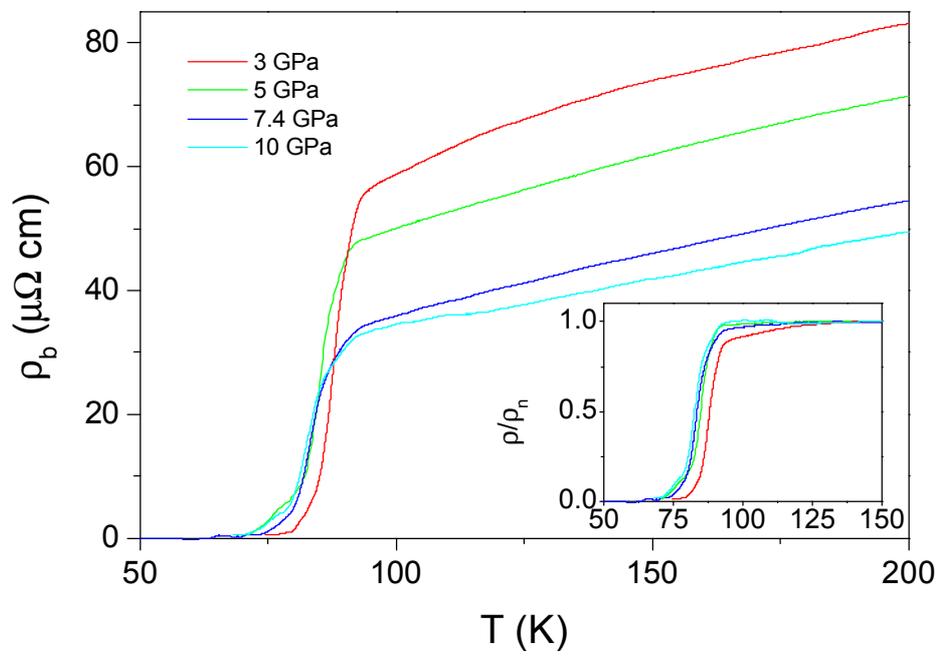

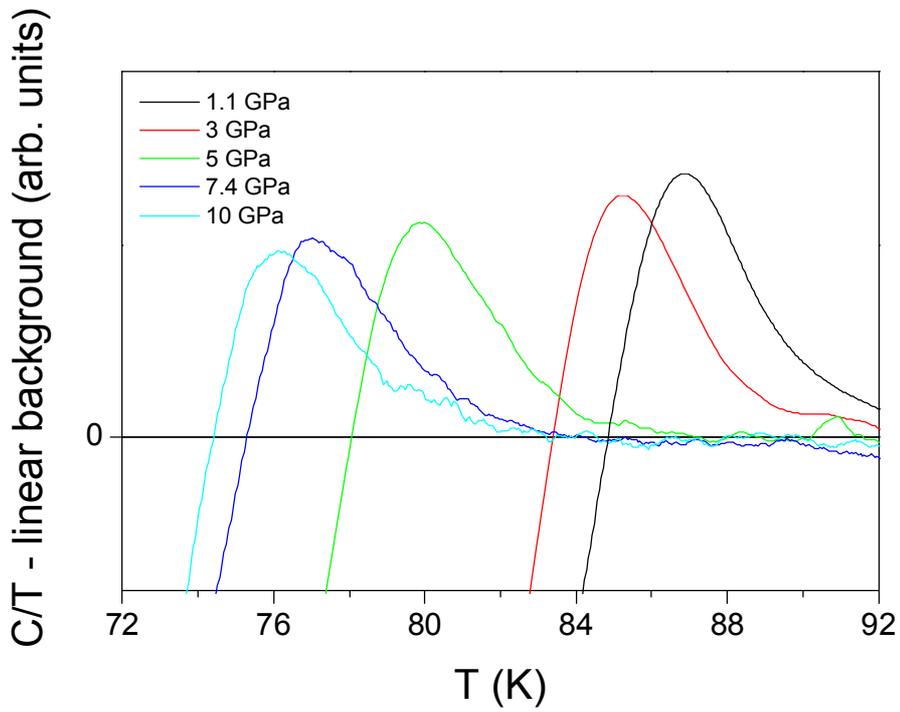

(a)

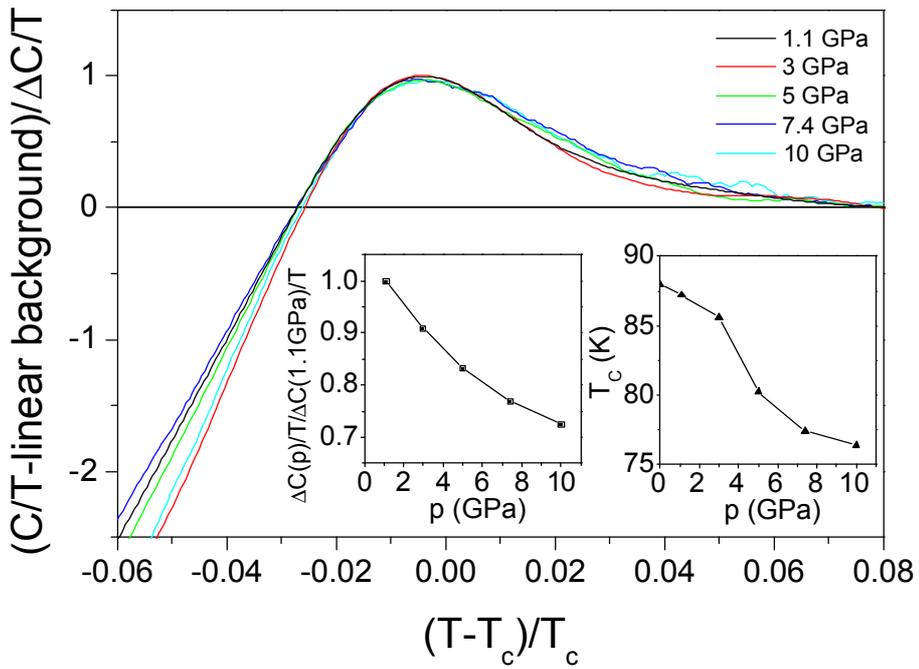

(b)

4a
4b

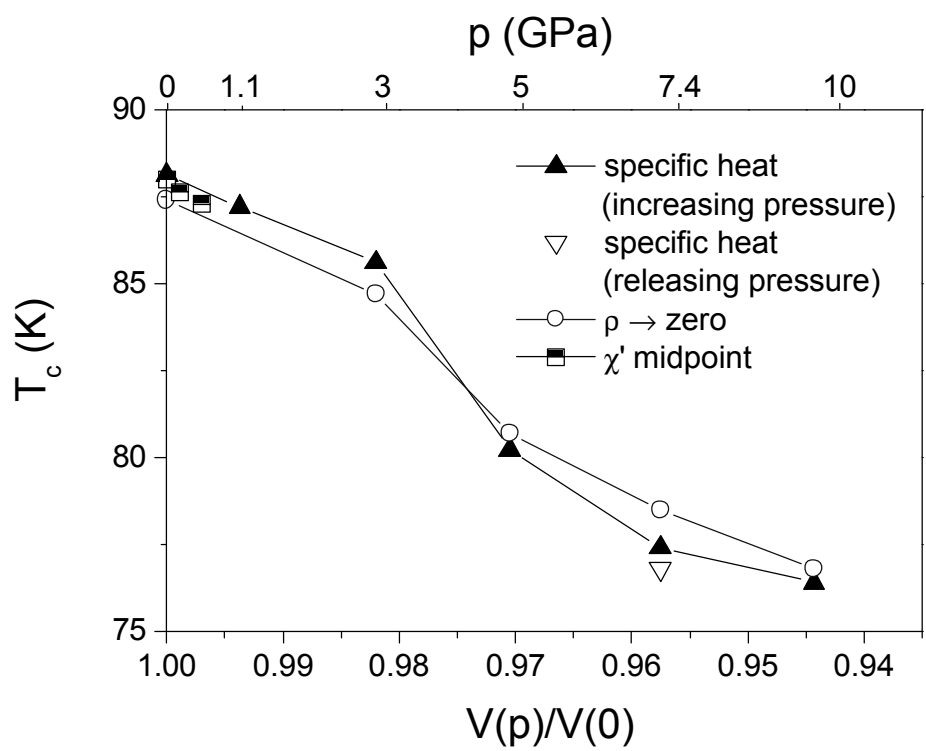

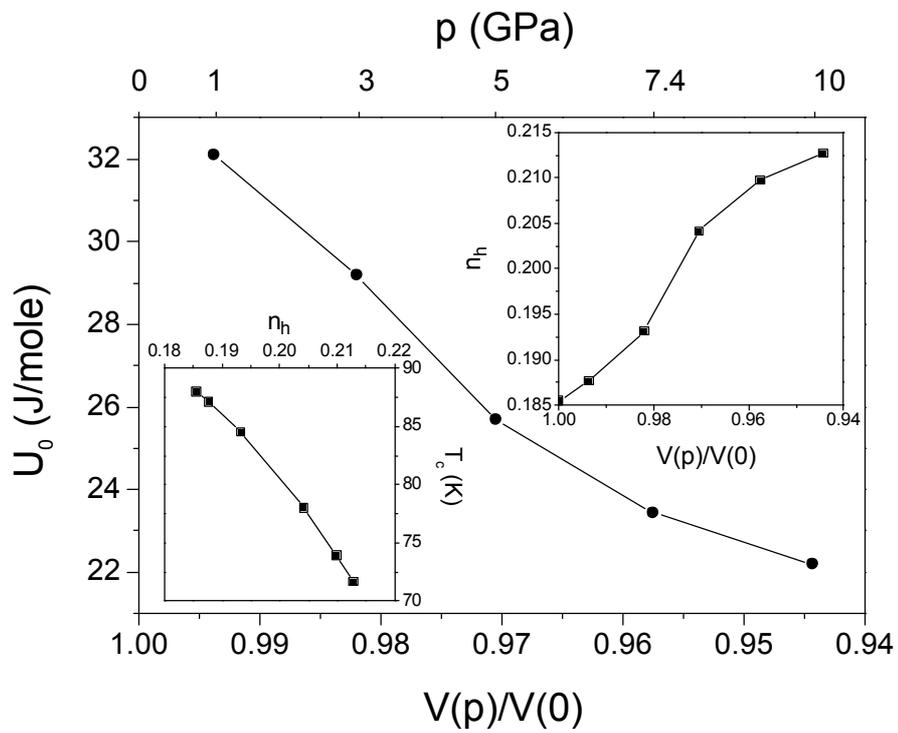

(a)

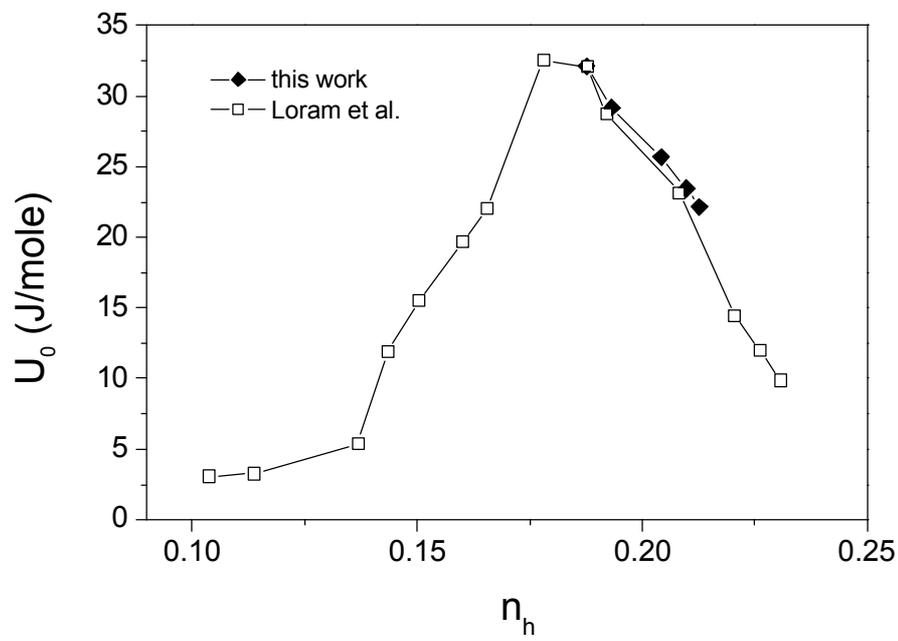

(b)

6a
6b